\documentclass{article}%
\usepackage{amsmath}
\usepackage{amsfonts}
\usepackage{amssymb}
\usepackage{mathrsfs}
\usepackage{graphicx}
\usepackage{bm}
\usepackage{cite}
\begin{document}

\title{Quantum collapse dynamics with attractive densities}
\author{F. Lalo\"{e} \footnote{laloe at lkb.ens.fr}\\LKB, ENS-Universit\'{e} PSL, CNRS, 24 rue Lhomond, 75005\ Paris, France}
\date{\today}
\maketitle

\begin{abstract}
We discuss a model of spontaneous collapse of the quantum state that does not require adding any stochastic processes to the standard dynamics. The additional ingredient with respect to the wave function is a position in the configuration space, which drives the collapse in a completely deterministic way. This new variable is equivalent to a set of positions of all the particles, i.e. a set of Bohmian positions, which obey the usual guiding equation of Bohmian theory. Any superposition of quantum states of a macroscopic object occupying different regions of space is projected by a localization process onto the region occupied by the positions. Since the Bohmian positions are well defined in a single realization of the experiment, a space localization into one region is produced. The mechanism is based on the correlations between these positions arising from the cohesive forces inside macroscopic objects.

The model introduces two collapse parameters, which play a very similar role to those of the GRW and CSL theories. With appropriate values of these parameters, we check that the corresponding dynamics rapidly projects
superpositions of macroscopic states localized in different regions of space into a single region, while is keeps a negligible effect in all situations where the predictions of standard quantum dynamics are known to be
correct. The possible relations with gravity are briefly speculated.  We then study the evolution of the density operator and a mean-field approximation of the dynamical equations
of this model, as well as the change of the evolution of the momentum
introduced by the localization process. Possible theoretical interpretations are finally discussed. Generally speaking, this model introduces a sharper border between the
quantum and classical world than the GRW and CSL theories, and leaves a broader range of acceptable values for the parameters.

\end{abstract}

\tableofcontents

\vspace{1cm}

\begin{center}
********
\end{center}

\vspace{1cm}

The standard linear Schr\"{o}dinger equation predicts the possible occurrence of quantum
superpositions of macroscopically distinguishable states
(QSMDS).\ This leads for instance to the famous Schr\"{o}dinger cat paradox
\cite{Schrodinger-cat, Trimmer}, to the so called measurement problem,
etc.\ Nevertheless, such QSMDS are apparently never observed, even in
experiments involving \textquotedblleft macroscopic quantum
phenomena\textquotedblright\ \cite{Leggett} such as superfluidity or
superconductivity.\ The problem arising from this apparent contradiction has given rise
to a huge literature \cite{measurement-problem-1, measurement-problem-2}.
Since Bohr, numerous authors have proposed various interpretations of quantum mechanics to deal with
this problem \cite{Laloe}.\ Another possible approach, nevertheless, is  to forbid the occurrence of
superpositions of QSMDS by modifying the
dynamics of quantum mechanics, for instance by adding a
small non-linear and stochastic term to the Schr\"{o}dinger equation. This is the basic
idea of \textquotedblleft spontaneous collapse theories\textquotedblright ,
such as the GRW \cite{GRW} and CSL \cite{CSL, CSL-2} theories; for reviews, see for
instance \cite{Bassi-Ghirardi} or \cite{Bassi-Lochan-Satin-Singh-Ulbricht}.
It has also been proposed to introduce collapse mechanisms that are driven by
gravity \cite{Diosi-1989, GGR-1990, Penrose-1996, Pearle-Squires}. A common feature of these
theories is the introduction of a stochastic term in the Schr\"{o}dinger dynamics.

Here we propose a model of spontaneous collapse where the dynamics is
deterministic -- as we will see, it is actually more a class of models than a specific model since, for instance, the localization
function can be chosen in several ways.\ Instead of adding stochastic processes or functions to the standard quantum description by a wave function, in the configuration space we add a position that drives the collapse mechanism. Adding positions to standard quantum mechanics is of course the basic idea of the de Broglie-Bohm (dBB)
interpretation \cite{de-Broglie, Bohm, Holland, Duerr-et-al, Bacchiagaluppi-Valentini, Bricmont}. Our model can therefore be seen as nothing but a combination of the dBB theory with spontaneous collapse theories.

A basic remark is that, in most cases, the number density of the Bohmian
positions in ordinary space coincides almost perfectly with the quantum
single-particle-density obtained from the many-body state vector; this is a
consequence of the so called \textquotedblleft quantum
equilibrium\textquotedblright\ condition, which in turn results from a dynamic that forces the Bohmian positions to follow
the wave function. Nevertheless, in Schr\"{o}dinger-cat like situations, or
after a quantum measurement has been performed, this is no longer the case: on
the one hand, because the quantum state splits into several macroscopically
distinct components, the quantum density divides into two (or more)
disconnected regions of space, corresponding for instance to different
positions of the pointer of the measurement apparatus; on the other hand, the
individual Bohmian positions of the particles must remain all clustered together in only one
of these regions. This clustering is a consequence of the internal cohesive
forces inside solid objects: the quantum Hamiltonian allows
quantum superpositions of states where all particles move together in one, or
another, region of space, but forbids states where some of the particles
are in one region and others in another region (the cohesive forces inside the pointer of the measurement apparatus forbid
states where the pointer is broken into two
parts). Because the set of all the Bohmian positions must define a point in
configuration space where the wave function does not vanish, they must remain
grouped together.

As a consequence, in one region of space (occupied for instance by the pointer
indicating a definite result), the single particle Bohmian density is much larger than the
density predicted by the quantum superposition; in another
region of space it is smaller, since the Bohmian density vanishes while the
quantum density does not. The basic idea of our model is to introduce a
dynamics where the quantum state vector is attracted to the first region, and
repelled from the others. The quantum dynamics
obtained in this way is completely deterministic: in a given
realization of an experiment, the only random element is the initial Bohmian
position of the configuration space of the physical system; once these
positions are determined, no random process takes place (as is also the case in dBB theory).

Our purpose is not, of course, to claim that the dynamics we propose is highly plausible. The main conceptual interest of such models is their very
existence, which proves that such approaches are neither impossible nor
contradictory with known experimental results.\ This is similar to the
existence of the dBB theory, which shows that some theorems concerning the
impossibility of additional quantum variables are irrelevant. In a previous
article \cite{SDAP}, we have already proposed a dynamics that also includes an
attraction of the state vector towards regions of high Bohmian
densities.\ Here we generalize and improve that model by introducing a spatial localization
term that introduces even smaller perturbations (except, of course, situations involving QSMDS), because the added
differential term in the dynamics remains almost zero in most cases. As a consequence, a larger flexibility is obtained for the values of the parameters of the dynamics; a relation with the Newton constant of gravity then becomes possible.

\section{Dynamic equation with a localization term}

We consider a system of $N$ identical spinless particles associated with a quantum
field operator $\Psi\left(  \mathbf{r}\right)  $, which is defined at each
point $\mathbf{r}$ of ordinary 3D space.

\subsection{Densities}

When the system is in state $\left\vert \Phi\right\rangle $, the local
(number) density $D_{\Phi}\left(  \mathbf{r}\right)  $ of particles at
$\mathbf{r}$ is:%
\begin{equation}
D_{\Phi}\left(  \mathbf{r}\right)  =\frac{\left\langle \Phi\right\vert
\Psi^{\dagger}\left(  \mathbf{r}\right)  \Psi\left(  \mathbf{r}\right)
\left\vert \Phi\right\rangle }{\left\langle \Phi\right.  \left\vert
\Phi\right\rangle } \label{sdap-1}%
\end{equation}
In dBB theory, the local density $D_{B}\left(  \mathbf{r}\right)  $ of Bohmian
positions is a sum of delta functions:%
\begin{equation}
D_{B}\left(  \mathbf{r,}t\right)  =\sum_{n=1}^{N}\delta\left(  \mathbf{r}%
-\mathbf{q}_{n}\right)  \label{sdap-2}%
\end{equation}
where the sum runs over all $N$ particles with Bohmian position $\mathbf{q}%
_{n}\left(  t\right)  $.

We wish to introduce a dynamics that favors evolutions where $D_{\Phi}\left(
\mathbf{r}\right)  $\ is attracted towards regions where $D_{B}\left(
\mathbf{r}\right)  $ takes higher values than $D_{\Phi}\left(  \mathbf{r}%
\right)  $, and repelled from regions where the opposite is true. Nevertheless, since $D_{B}\left(  \mathbf{r}\right)  $\ is singular, it is
useful to introduce a space average. For this
purpose, we choose a distance $a_{L}$ and a function $A(\mathbf{r)}$ that is
localized around the origin of space within a distance $a_{L}$, for instance:%
\begin{equation}
A_{L}(\mathbf{r)} = e^{-\left(  \mathbf{r}-\mathbf{r}^{\prime}\right)
^{2}/\left(  a_{L}\right)  ^{2}}\label{sdap-2-bis}%
\end{equation}
We then introduce the following integrals of $D_{\Phi}$ and $D_{B}$:%
\begin{equation}
N_{\Phi}\left(  \mathbf{r},t\right)  =\int\text{d}^{3}r^{\prime}%
~A_{L}(\mathbf{r}-\mathbf{r}^{\prime}\mathbf{)}~D_{\Phi}\left(  \mathbf{r}%
^{\prime},t\right)  \label{sdap-3}%
\end{equation}
and:%
\begin{equation}
N_{B}\left(  \mathbf{r,}t\right)  =\int\text{d}^{3}r^{\prime}~A_{L}%
(\mathbf{r}-\mathbf{r}^{\prime}\mathbf{)}~D_{B}\left(  \mathbf{r}^{\prime
},t\right)  =\sum_{n=1}^{N}~A_{L}(\mathbf{r}-\mathbf{q}_{n}\mathbf{)}%
~\label{sdap-5}%
\end{equation}

The Gaussian form (\ref{sdap-2-bis}) is one possibility, but we could also
have made different choices, for instance:%
\begin{equation}
A_{L}(\mathbf{r)=}\frac{\left(  a_{L}\right)  ^{s}}{\left(  a_{L}\right)
^{s}+\left(  \mathbf{r}\right)  ^{s}}\label{sdap-5-bis}%
\end{equation}
where $s$ is any integer number that is larger than $2$. All the discussion of this
article is actually independent of a particular choice of the localization
function.

The order of magnitude of $N_{\Phi}\left(  \mathbf{r},t\right)  $
is the (quantum) average number of particles within a volume $\left(
a_{L}\right)  ^{3}$ around point $\mathbf{r}$; similarly, the order of
magnitude of $N_{B}\left(  \mathbf{r,}t\right)  $ is the number of Bohmian
positions inside the same volume.\ We have $0\leq$ $N_{\Phi}\left(  \mathbf{r}%
,t\right)  ,N_{B}\left(  \mathbf{r,}t\right)  $. Since:
\begin{equation}
\int\text{d}^{3}r~N_{\Phi}\left(  \mathbf{r},t\right)  =\int\text{d}%
^{3}r~N_{B}\left(  \mathbf{r},t\right)  = N \int\text{d}^{3}r ~  A_{L}(\mathbf{r)}  \label{sdap-6}%
\end{equation}
both these numbers have
an upper bound that is $N$ times the space integral of $A_L (\mathbf r)$..

\subsection{Attractive dynamics}

We then define the (dimensionless) localization operator $L\left(  t\right)  $
by:%
\begin{equation}
L\left(  t\right)  =\int\text{d}^{3}r~\Delta\left(  \mathbf{r},t\right)
~\Psi^{\dagger}\left(  \mathbf{r}\right)  \Psi\left(  \mathbf{r}\right)
\label{sdap-8}%
\end{equation}
where $\Delta(\mathbf{r},t)$ is defined as the difference:%
\begin{equation}
\Delta(\mathbf{r},t)=N_{B}\left(  \mathbf{r},t\right)  -N_{\Phi}\left(
\mathbf{r},t\right)  \label{sdap-7}%
\end{equation}
This allows us to introduce a dynamics that favors evolutions where $D_{\Phi}\left(
\mathbf{r}\right)  $\ is attracted towards regions where $\Delta\left(
\mathbf{r},t\right)  >$ $0$.\ For this purpose, we add to the usual
Hamiltonian $H\left(  t\right)  $ a localization term that is proportional to
$L\left(  t\right)  $, and write the modified Schr\"{o}dinger equation:%
\begin{equation}
i\hslash\frac{d}{dt}\left\vert \Phi\left(  t\right)  \right\rangle =\Big[
H\left(  t\right)  +i\hslash\gamma_{L}~L\left(  t\right)  \Big]  \left\vert
\Phi\left(  t\right)  \right\rangle \label{sdap-9}%
\end{equation}
where $\gamma_{L}$ is a constant localization rate. The new term in the Hamiltonian increases the modulus of the wave
function in regions where $\Delta\left(  \mathbf{r},t\right)  $ is positive,
reduces it in regions where the opposite is true.

Relation (\ref{sdap-6}) implies that the space integral of $\Delta\left(
\mathbf{r},t\right)  $ vanishes:
\begin{subequations}
\begin{equation}
\int\text{d}^{3}r~\Delta\left(  \mathbf{r},t\right)  =0 \label{sdap-7-bis}%
\end{equation}
But, as in \cite{SDAP}, we could also have defined $\Delta\left(
\mathbf{r},t\right)  $ as:
\begin{equation}
\Delta(\mathbf{r},t)=N_{B}\left(  \mathbf{r},t\right)  \label{sdap-7b}%
\end{equation}
The right hand side of relation (\ref{sdap-7-bis}) would then be given by \ref{sdap-5}.

Since the operator acting in the right-hand side of (\ref{sdap-9}) is not
Hermitian, this equation of evolution does not conserve the norm of
$\left\vert \Phi\right\rangle $. Nevertheless, if desired, one can easily
obtain a normalized state vector $\left\vert \overline{\Phi}\right\rangle $:%
\end{subequations}
\begin{equation}
\left\vert \overline{\Phi}\right\rangle =\left\vert \overline{\Phi}\left(
t\right)  \right\rangle =\frac{1}{\sqrt{\left\langle \Phi(t)\right.
\left\vert \Phi(t)\right\rangle }}\left\vert \Phi(t)\right\rangle
\label{sdap-10}%
\end{equation}
which obeys the following equation of evolution:%
\begin{align}
\frac{\text{d}}{\text{d}t}\left\vert \overline{\Phi}\left(  t\right)
\right\rangle  &  =\frac{1}{\sqrt{\left\langle \Phi(t)\right.  \left\vert
\Phi(t)\right\rangle }}\frac{\text{d}}{\text{d}t}\left\vert \Phi
(t)\right\rangle -\frac{1}{2\left\langle \Phi(t)\right.  \left\vert
\Phi(t)\right\rangle ^{3/2}}\left(  \frac{\text{d}}{\text{d}t}\left\langle
\Phi(t)\right.  \left\vert \Phi(t)\right\rangle \right)  \left\vert
\Phi(t)\right\rangle \nonumber\\
&  =\frac{1}{i\hslash}\left[  H\left(  t\right)  +i\hslash\gamma_{L}~L\left(
t\right)  \right]  \left\vert \overline{\Phi}\left(  t\right)  \right\rangle
-\gamma_{L}\frac{1}{\left\langle \Phi(t)\right.  \left\vert \Phi
(t)\right\rangle ^{3/2}}~\left\langle \Phi(t)\right\vert L\left(  t\right)
\left\vert \Phi(t)\right\rangle ~\left\vert \Phi(t)\right\rangle \nonumber\\
&  =\frac{1}{i\hslash}\left[  H\left(  t\right)  +i\hslash\gamma_{L}~L\left(
t\right)  \right]  \left\vert \overline{\Phi}\left(  t\right)  \right\rangle
-\gamma_{L}\int\text{d}^{3}r~\Delta\left(  \mathbf{r},t\right)  ~D_{\Phi
}\left(  \mathbf{r}\right)  ~\left\vert \overline{\Phi}\left(  t\right)
\right\rangle \label{sdap-11}%
\end{align}
We then obtain:%
\begin{equation}
i\hslash\frac{\text{d}}{\text{d}t}\left\vert \overline{\Phi}\left(  t\right)
\right\rangle =\left[  H\left(  t\right)  +i\hslash\gamma_{L}~\overline
{L}\left(  t\right)  \right]  \left\vert \overline{\Phi}\left(  t\right)
\right\rangle \label{sdap-12}%
\end{equation}
with:%
\begin{equation}
\overline{L}\left(  t\right)  =\int\text{d}^{3}r~\left[  \Psi^{\dagger}\left(
\mathbf{r}\right)  \Psi\left(  \mathbf{r}\right)  -D_{\Phi}\left(
\mathbf{r}\right)  \right]  \Delta\left(  \mathbf{r,}t\right)  \label{sdap-13}%
\end{equation}

If $\Delta\left(  \mathbf{r,}t\right)  $ is a constant in space, we remark
that the effect of $\overline{L}\left(  t\right)  $ on any ket with a fixed
number of particles vanishes.\ This is of course the case if $a_{L}=0$
($N_{B}$, $N_{\Phi}$ and $\Delta$ then vanish), but also if $a_{L}=\infty$
(then $N_{B}=N_{\Phi}$ and $\Delta\left(  \mathbf{r,}t\right)$ vanishes again). The localization term is
effective only if $a_{L}$ takes an intermediate, finite, value for which
$\Delta\left(  \mathbf{r,}t\right)  $ varies in space.

\subsection{Modified Schr\"{o}dinger equation}
\label{effect-localization}

We now study the modified Schr\"{o}dinger equation (\ref{sdap-12}) in the
position representation.
The localization operator (\ref{sdap-13}) contains a first term in
$\Psi^{\dagger}\left(  \mathbf{r}\right)  \Psi\left(  \mathbf{r}\right)  $
that has the form of a (symmetric) potential operator, diagonal in the position
representation.\ This operator can also be written as a summation
over all particles \cite{vol-3}:%
\begin{equation}
\int\text{d}^{3}r~\left[  \Psi^{\dagger}\left(  \mathbf{r}\right)  \Psi\left(
\mathbf{r}\right)  \right]  \Delta\left(  \mathbf{r,}t\right)  =\sum_{n=1}%
^{N}\Delta\left(  \mathbf{R}_{n}\mathbf{,}t\right)  \label{sdap-14}%
\end{equation}
where $\mathbf{R}_{n}$ is the position operator associated with the position
of particle $n$. We therefore have:%
\begin{equation}
\left\langle 1:\mathbf{r}_{1};2:\mathbf{r}_{2};..~;N:\mathbf{r}_{N}\right\vert
\int\text{d}^{3}r~\Psi^{\dagger}\left(  \mathbf{r}\right)  \Psi\left(
\mathbf{r}\right)  ~\Delta\left(  \mathbf{r,}t\right)  ~\left\vert
\overline{\Phi}\left(  t\right)  \right\rangle =\sum_{n=1}^{N}\Delta\left(
\mathbf{r}_{n}\mathbf{,}t\right)  ~\overline{\Phi}\left(  \mathbf{r}%
_{1},\mathbf{r}_{2},..,\mathbf{r}_{n},..~,\mathbf{r}_{N};t\right)
\label{calc-2}%
\end{equation}
where $\overline{\Phi}\left(  \mathbf{r}_{1},\mathbf{r}_{2},..,\mathbf{r}%
_{n},..\mathbf{r}_{N};t\right)  $ is the wave function representing the $N$
particle system in configuration space (for the sake of simplicity, we assume
that the particles have no spin).

As for the second term in the right hand side of (\ref{sdap-13}), it is just a
c-number, proportional to the constant $<\Delta>$ defined by:%
\begin{equation}
<\Delta>~=\frac{1}{N}\int\text{d}^{3}r~D_{\Phi}\left(  \mathbf{r}\right)
~\Delta\left(  \mathbf{r,}t\right)  \label{calc-3}%
\end{equation}
Since $D_{\Phi}\left(  \mathbf{r}\right)  /N$ is a distribution over space that is
normalized to unity, $<\Delta>$ is the average of $\Delta\left(
\mathbf{r,}t\right)  $ over the one-body density of the wave function.

The evolution of the wave function due to the localization term is therefore:
\begin{equation}
\left.  \frac{\text{d}}{\text{d}t}\right\vert _{\text{loc}}\overline{\Phi
}\left(  \mathbf{r}_{1},\mathbf{r}_{2},..~,\mathbf{r}_{n},..,\mathbf{r}%
_{N};t\right)  =\gamma_{L}\left[  \sum_{n=1}^{N}\Delta\left(  \mathbf{r}%
_{n}\mathbf{,}t\right)  -N<\Delta>\right]  \overline{\Phi}\left(
\mathbf{r}_{1},\mathbf{r}_{2},..~,\mathbf{r}_{n},..,\mathbf{r}_{N};t\right)
\label{calc-4}%
\end{equation}
The wave function tends to increase in regions where many $\Delta\left(
\mathbf{r}_{n}\mathbf{,}t\right)  $ are larger than their space average value
$<\Delta>$, and tends to decrease in regions of space where the opposite is true.

\subsection{Coupled evolutions}
\label{coupled-evolutions}

We assume that the Bohmian positions $\mathbf{q}_{n}$ evolve
according to the usual Bohmian equation of motion:%
\begin{equation}
\frac{d\mathbf{q}_{n}\left(  t\right)  }{dt}=\hslash\frac{\bm\nabla_{n}\xi}{m} \label{sdap-5-ter}%
\end{equation}
where $\xi\left(  \mathbf{r}_{1},\mathbf{r}_{2},..,\mathbf{r}_{N}\right)  $ is
the phase of the wave function $\Phi\left(  \mathbf{r}_{1},\mathbf{r}%
_{2},..,\mathbf{r}_{N}\right)  $, and $\bm\nabla_{n}$
the gradient taken with respect to $\mathbf{q}_{n}$. In standard dBB theory,
this relation ensures that the condition of \textquotedblleft quantum
equilibrium\textquotedblright\  is satisfied at any time, if it is satisfied at the
initial time.\ Nevertheless, this assumes that the equation of
evolution of the wave function is the standard Schr\"{o}dinger equation, which
is no longer the case in our model.\ In \cite{SDAP}, we argued that this was
not a serious problem, since the localization term is very small, while Towler, Russell and Valentini \cite{Valentini-2005, Valentini-2012} have shown that a fast
relaxation process drives the system quickly back to quantum equilibrium.\ Assuming quantum equilibrium should therefore still be an
excellent approximation in most cases; we come back to this point in more detail in \S~\ref{quantum-equilibrium}.

\section{Collapse dynamics and  time constants, examples}
\label{examples}

We now explore the predictions of the model for different choices of the
constants.\ We will see that the model is very robust:  even with a large variation of
its constants, it
remains compatible with experimental observations. To illustrate this point, we choose either of the following pair of values:
\begin{subequations}
\label{calc-112-other}%
\begin{align}
\gamma_{L}  &  =10^{-24}~\text{s}^{-1}\label{calc-112a-other}\\
a_{L}  &  =1\mu \text{m} =10^{-6}~\text{m} \label{calc-112b-other} %
\end{align}
or the same values as those of GRW\ and CSL:
\end{subequations}
\begin{subequations}
\label{calc-112}%
\begin{align}
\gamma_{L}  &  =10^{-16}~\text{s}^{-1}\label{calc-112a}\\
a_{L}  &  =10~\mu \text{m} =10^{-5}~\text{m} \label{calc-112b}%
\end{align}
We now examine a few situations where the consequences of these choices can be evaluated, depending on whether a superposition of macroscopically distinct states is involved or not.

\subsection{Various situations}
\label{various-situations}

In (\ref{sdap-13}), only the first term in the bracket defining the localization operator $\overline{L}\left(  t\right)
$ is physically effective.\ This is
because the second term in $D_{\Phi}\left(  \mathbf{r}\right)  $ introduces
only a change of the norm of the whole state vector, which does not change its
physical content; we will therefore ignore it in this section. Moreover, if
$\Delta\left(  \mathbf{r},t\right)  $ is uniform in space, condition
(\ref{sdap-7-bis}) shows that $\overline{L}\left(  t\right)  $ vanishes: the
localization operator is non-zero only if, after averaging over a volume $(a_{L})^{3}$,
the Bohmian and quantum densities
still have different variations in space.\ Consider a region of space $R$ where the difference $N_{B}\left(
\mathbf{r},t\right)  -N_{\Phi}\left(  \mathbf{r},t\right)  $ takes a typical
value $\Delta_{R}$; a localization process occurs in this region with a rate
$\gamma_{L}\Delta_{R}N_{R}$, where $N_{R}$ is the number of particles
contained in $R$.\ A distortion of the wave function, and therefore a physical
modification of the properties of the system, occurs only if two (or more)
regions $R$ and $R^{\prime}$ have values for $\Delta_{R}$ and
$\Delta_{R^{\prime}}$  with different values.

For microscopic systems, the effect of the localization term remains extremely
slow.\ Since both $\Delta_{R}$ and $N_{R}$ have an upper bound equal to the
number of particles $N$, no localization in any region can occur at a rate
higher than $\gamma_{L}N^{2}$. If we assume for instance $N=10^{6}$, with both
choices (\ref{calc-112-other}) and (\ref{calc-112}) we obtain upper bounds of
the localization rate of the order of $10^{-4}$ s$^{-1}$, or even much less.
Moreover, these upper bounds can be approached only if the sign of
$\Delta\left(  \mathbf{r},t\right)  $ is opposite in two different regions of
space extending over more that $a_{L}$: if a microscopic system is localized
in space in a region smaller than $a_{L}$, its localization rate is therefore even much smaller.

Clearly, to obtain a significative localization effect, it is required to have
at the same time a large number of particles involved and spatial separations
exceeding $a_{L}$. The optimal situation to detect an effect with a
microscopic or mesoscopic system would probably be an interference experiment with a very
large molecule or cluster \cite{Arndt-et-coll}, assuming that the distance
between the slits is larger than $a_{L}$. Nevertheless, even with $10^{8}$
particles in the cluster, the time of flight along the two different paths
should be at last $1$ second for a significant localization effect to be
obtained if (\ref{calc-112a}) is selected, more than one year if
(\ref{calc-112a-other}) is selected. In most cases, mesoscopic systems seem to be unaffected by the localization term.

The situation is radically different if a QSMDS is created, as is indeed the
case during a quantum measurement. We assume that the distance between the
positions of the pointer of the measurement apparatus indicating different
results (or the distance between the positions of macroscopically distinct
states) is larger than $a_{L}$.$\ $For simplicity, we assume that only two
positions are possible, corresponding to two results of measurement (the generalization to
more results is trivial).\ In
this case, $\Delta\left(  \mathbf{r},t\right)  $ has large but opposite values
in two distant regions of space. This is because $N_{\phi}\left(
\mathbf{r,}t\right)  $ and $N_{B}(\mathbf{r},t\mathbf{)}$ have completely
different behaviors: on the one hand, the quantum density of
particles and $N_{\phi}\left(  \mathbf{r,}t\right)  $ is distributed among two
wave packets; on the other hand, in a single realization of the experiment,
the Bohmian factor $N_{B}(\mathbf{r},t\mathbf{)}$ vanishes in one of the wave
packets (the empty wave packet), while it takes a maximum value in the
other. As discussed in the introduction, this results from the cohesive forces inside the pointer, which create
strong quantum correlations between the positions of its constituent
particles: in quantum mechanics, these particles can be at the same time in
two different regions of space, but they have to remain all together in the
same region.\ Therefore, since the Bohmian positions must define a point in
configuration space where the $N$-particle wave function does not vanish, they
have to remain grouped: all of them are in the same wave packet, none is in
the other (empty wave). In one channel, $\Delta(\mathbf{r},t)\simeq
N_{B}(\mathbf{r},t\mathbf{)-}N_{\Phi}(\mathbf{r},t\mathbf{)\simeq+}N_{\Phi
}(\mathbf{r},t)$, in the other channel $\Delta(\mathbf{r},t)\simeq
0\mathbf{-}N_{\Phi}(\mathbf{r},t\mathbf{)\simeq-}N_{\Phi}(\mathbf{r},t)$.

In a quantum measurement situation, we can for instance assume that the
apparatus contains a pointer that is a solid containing $N_{\Phi}\simeq
10^{11}$ particles per cubic micron (this rough order of magnitude seems to be
reasonable for the number of atoms; the number of nucleons, or electrons, would
be larger). If the total number of particles in the pointer is $N_{P}$, the
differential rate of relaxation between the full wave (that associated with
the result of measurement) and the other empty wave(s) is of the order of
$2\gamma N_{\Phi}N_{P}$.\ If for instance the pointer is a tiny cube with
$100\mu$m side only, $N_{P}\simeq10^{17}$, and we see that a superposition of
two (or more) spatially separate states of the pointer disappears in about
$10^{-4}$s. with (\ref{calc-112-other}), or $10^{-12}$ s. with
(\ref{calc-112}).\ This is the time it takes the measurement apparatus to
display a definite result.\ The model therefore ensures a rapid collapse of
the wave function, even for tiny apparatuses of measurement. Another remark is that the so called \textquotedblleft surrealistic trajectories\textquotedblright should not exist within this model; actually, even within standard dBB theory, they already do not occur with macroscopic bodies \cite{Tastevin-Laloe}.

We also note that, even if $N_{\phi}\left(  \mathbf{r,}t\right)  $ may strongly differ
from $N_{B}(\mathbf{r},t\mathbf{)}$ during a single realization of the
experiment, this is only a short transient effect taking place while the
measurement is completed; then the additional localization term in the
Schr\"{o}dinger equation rapidly ensures that $N_{\phi}\left(  \mathbf{r,}%
t\right)  $ relaxes towards $N_{B}(\mathbf{r},t\mathbf{)}$. It modifies the
state vector so that $N_{\phi}\left(  \mathbf{r,}t\right)  $ vanishes in all
wave packets but one. After the measurement is completed, the dynamical relaxation process studied in
\cite{Valentini-2005, Valentini-2012} ensures that the difference $N_{\phi}\left(  \mathbf{r,}%
t\right)  -N_{B}(\mathbf{r},t\mathbf{)}$ tends rapidly to zero. We discuss in \S~\ref{quantum-equilibrium} why it is possible to assume that the quantum equilibrium condition is restored when a second
experiment is started.

One may wonder if the addition of a nonlinear localization term in the dynamics could produce dramatic unexpected effects, despite the extremely small value of the nonlinear coefficient. Indeed, the very purpose of the model is to obtain a dramatic effect during a measurement process: the state vector is suddenly projected onto one of its components, and all empty waves disappear. Does this extend to other situations? Mathematically, a similar question occurs with the Gross-Pitaevskii equation, which describes interacting Bose-Einstein condensates within mean field theory: an ideal gas is only marginally stable, since an infinitesimal attractive nonlinear term is sufficient to produce a collapse of the boson gas. The question then is: in what circumstances is the standard dBB theory only marginally stable with respect to the addition of an arbitrarily small nonlinear perturbation? Within our model, strong localizations effects occur as soon as the Bohmian density differs significantly from the quantum density. In the abscence of a QSMDS states, there is no special reason why this difference should be large. In a localized piece of bulk solid for instance, the Bohmian positions of the particles are randomly distributed in the volume occupied by the solid, and the coincidence between the two densities is rather good; the two terms in (\ref{calc-4}) then almost cancel each other (moreover, even without this cancellation, the localization term would only then to localize the solid inside its own volume, with no dramatic effect). The localization term is efficient mostly in the presence of quantum superpositions where many particles occupy different regions of space, that is basically QSMDS.

This discussion of various possible physical situations shows that a broad class of models is indeed compatible with the experimental observations that are known at present. In fact, we can choose either definition (\ref{sdap-7}) or (\ref{sdap-7b}) of $\Delta\left(\mathbf{r},t\right) $, and then select either (\ref{calc-112-other}) or (\ref{calc-112}) for the constants of the model: in all cases we obtain a fast time constant for the appearance of a single result of measurement, without introducing appreciable perturbations of microscopic systems. The basic reason for this flexibility is the quadratic dependence of the product $N_{\Phi}N_{P}$ in the density of macroscopic objects, which introduces a fast relaxation rate even with very small values of $\gamma_L$. In other words, for macroscopic objects, the relaxation rate varies proportionally to the square of the Avogadro number, which is an enormous number.

\subsection{Quantum equilibrium}
\label{quantum-equilibrium}

The quantum equilibrium condition applies in configuration space, and therefore introduces more stringent condition than the  equality of densities in ordinary
space discussed in the preceding subsection.\ It also relates to an ensemble of
realizations of the same experiment. Assuming that, initially,  the Bohmian positions are randomly distributed, and that their distribution coincides with the
quantum probability density (the modulus square of the wave function in the configuration space), the usual dBB theory ensures that  the
coincidence remains exact at all times.

This condition cannot be directly transposed to our modified dynamics, where different Bohmian positions associated with the same initial wave function lead to different wave functions at later times. The distribution of the positions can then no longer be compared to a single quantum probability density. For instance, just after a
measurement has been performed (or, more generally, when a QSMDS has been
projected onto one of its macroscopic components), the physical system is
described by several different wave functions, depending on the result obtained in
the experiment. We therefore have to modify the condition by requesting that,
if we consider only the sub-ensemble of realizations having provided a specific result of measurement, the new wave function provides a density
in configuration space that matches the distribution of Bohmian
positions for this sub-ensemble.\ Clearly, this condition cannot be exactly fulfilled at all times,
in particular during the (very short) projection process of the wave
function.\

Nevertheless, once the
projection process is complete, in the many particle system made of the entangled
measured system $S$ and measurement apparatus $M$, one can again rely on the dynamical relaxation
process discussed by Valentini et al. \cite{Valentini-2005, Valentini-2012} to restore quantum equilibrium.\ Indeed, these authors have shown that, at least  in simple systems, a statistical distribution of the Bohmian position relaxes very quickly towards the modulus square of the wave function.
Admittedly, we are making some extrapolation at this stage: as far as we know, there exist no
systematic study  in configuration
space of the Bohmian dynamics of large entangled systems.\ Valentini has nevertheless shown \cite{Valentini-1991} that the approach to quantum equilibrium can be derived from a statistical \textquotedblleft subquantum theorem\textquotedblright , based on statistical assumptions that are analogous to those of classical statistical mechanics. The assumption we are making at this stage could probably be tested by more systematic numerical simulations of
the Bohmian dynamics.

Within this scenario, assume that a second measurement is performed on the same
system $S$ after the first experiment has been completed (its
result has been registered). It is then legitimate to assume that the quantum
equilibrium is obeyed for the sub-ensemble of experiments that have given a
specific result in the first experiment.\ We then recover the standard rules of dBB mechanics and the Born rule, within a possible relative error of $10 ^{-16}$ or less if the whole experiment lasts one second; such an error rate is of course totally undetectable.

In other words, within our model, quantum equilibrium is no longer considered as a condition that is exactly met at all times. For instance, equilibrium is not yet reached while the result of an experiment is appearing on the pointer of a measurement apparatus; one has to wait until the result of measurement is fully registered. Quantum equilibrium is then rather seen as an emergent phenomenon \cite{Valentini-2009} that reappears after each measurement, so that the initial conditions for the next experiment that are extremely close to this equilibrium.

\subsection{Introducing the gravitational constant}
\label{gravitational}

Instead of two arbitrary constants as in (\ref{calc-112}), it is possible to
introduce only a single constant, for instance $a_{L}$, provided the Newton constant $G$ is taken into account. We may assume that:%
\end{subequations}
\begin{equation}
\gamma_{L}=\frac{m^{2}G}{\hbar a_{L}}\label{calc-116}%
\end{equation}
where $m$ is the mass of a nucleon for instance. If we choose value
(\ref{calc-112b-other}) for $a_{L}$, we obtain:%
\begin{equation}
\gamma_{L}\simeq\frac{10^{-54}~6~10^{-11}}{10^{-34}~10^{-6}}\simeq
10^{-24}~\text{s}^{-1}\label{calc-117}%
\end{equation}
which is indeed compatible with (\ref{calc-112a-other}). One can then assume
that the limit between the macroscopic and microscopic world occurs at a
characteristic length $a_{L}=1\mu$m, and then use relation (\ref{calc-116}) to
\textquotedblleft explain\textquotedblright\ why $\gamma_{L}$ has the
extremely small value given in (\ref{calc-112a-other}).

We can for instance consider that the collapsing field originates from the average
gravitational attraction of the other identical particles within a range
$a_{L}$. The source of this classical field is the average Bohmian
density (not the average quantum density), which localizes the quantum state in space. This is similar to the theory proposed in Ref. \cite{Pearle-Squires}, within a stochastic dynamics. In
quantum cosmogenesis, similar ideas have been proposed in Refs.
\cite{Peter-Pinho-et-al} \cite{Pinho-Pinto} within the dBB theory, in order to treat the metric of
general relativity classically by considering the Bohmian positions as the
sources of gravity.

Needless to say, in relations (\ref{calc-112-other}), we can multiply
$\gamma_{L}$ by any factor $\lambda$ and $a_{L}$ by $1/\lambda$ without
changing this agreement. Actually, relation (\ref{calc-116}) only determines a velocity
$c_{L}$ as:
\begin{equation}
c_{L}=\gamma_{L}a_{L}=\frac{m^{2}G}{\hbar} \label{calc-166-bis}%
\end{equation}
One can also generalize (\ref{calc-116}) by introducing a universal
dimensionless constant $\alpha_{L}$ as:%
\begin{equation}
\alpha_{L}=\frac{m^{2}G}{\hbar c_{L}} \label{calc-166-ter}%
\end{equation}
and consider models where this constant takes on any arbitrary
dimensionless value, for instance $2\pi$, $1/137$, etc.

\section{Density operator}
\label{density-operator}

We now examine the effect of the localization term on the evolution of the
density operator, either in a single realization of the experiment, or by
average over many realizations.

\subsection{Time evolution (single realization)}

If the system is in a normalized pure state $\left\vert \Phi\left(  t\right)
\right\rangle $, the density operator $\rho(t)$ is defined as:%
\begin{equation}
\rho(t)=\left\vert \Phi\left(  t\right)  \right\rangle \left\langle
\Phi\left(  t\right)  \right\vert \label{calc-118}%
\end{equation}
The density $D_{\Phi}\left(  \mathbf{r}\right)  $ is now defined by:%
\begin{equation}
D_{\rho}(\mathbf{r},t)=\text{Tr}\left\{  \Psi^{\dagger}\left(  \mathbf{r}%
\right)  \Psi\left(  \mathbf{r}\right) \, \rho(t)\right\}  \label{calc-120}%
\end{equation}
$N_{\Phi}\left(  \mathbf{r},t\right)  $ is replaced by $N_{\rho}\left(
\mathbf{r},t\right)  $, obtained by substituting $D_{\rho}(\mathbf{r},t)$ to
$D_{\Phi}\left(  \mathbf{r}\right)  $ in (\ref{sdap-3}). The same
changes are made in (\ref{sdap-7}) and in the definition (\ref{sdap-13}) of $\overline{L}\left(
t\right)  $. Then $\rho(t)$ evolves according to
the equation:%
\begin{equation}
i\hbar\frac{\text{d}}{\text{d}t}\rho(t)=\left[  H\left(  t\right)
,\rho(t)\right]  +i\hslash\gamma_{L}~\left[  \overline{L}\left(  t\right)
,\rho(t)\right]  _{+} \label{calc-119}%
\end{equation}
where $\left[  A,B\right]  _{+}$ is the anticommutator $AB+BA$ of the two
operators $A$ and $B$. This equation is nonlinear
since $D_{\rho}(\mathbf{r},t)$, and therefore $\overline{L}\left(  t\right)
$, depends on $\rho(t)$.

We check that:%
\begin{align}
i\hbar\frac{\text{d}}{\text{d}t}\text{Tr}\left\{  \rho(t)\right\}   &
=2i\hslash\gamma_{L}\text{Tr}\left\{  \int\text{d}^{3}r~\left[  \Psi^{\dagger
}\left(  \mathbf{r}\right)  \Psi\left(  \mathbf{r}\right)  -D_{\rho}\left(
\mathbf{r}\right)  \right]  \Delta\left(  \mathbf{r,}t\right)  ~\rho
(t)\right\} \nonumber\\
&  =2i\hslash\gamma_{L}\text{Tr}\left\{  \int\text{d}^{3}r~\left[  D_{\rho
}\left(  \mathbf{r}\right)  -D_{\rho}\left(  \mathbf{r}\right)  \right]
\Delta\left(  \mathbf{r,}t\right)  ~\rho(t)\right\}  =0 \label{calc-120-2}%
\end{align}

\subsection{Average over many realizations}

The evolution of the density operator describing the average of many
realizations of an experiment is given by the average of equation
(\ref{calc-119}) over these realizations.\ If the system contains a single
particle, its Bohmian position is different for each realization; during time evolution, it
explores various regions of the wave function, as shown in the figures of
\cite{Valentini-2005, Valentini-2012}.\ Therefore, when the average over many realizations is
taken, $N_{B}\left(  \mathbf{r,}t\right)  $ as well as $\Delta\left(
\mathbf{r,}t\right)  $ play the role of random functions.\ It the system
contains $N$ particles, $\Delta(\mathbf{r},t)$ is then the sum of $N$
fluctuating functions.\ In both cases, (\ref{calc-119}) becomes similar to a stochastic
differential equation.\ We remark that, if the initial distribution of Bohmian
variables coincides with the quantum distribution, the ensemble average of
$\Delta(\mathbf{r},t)$ vanishes.\ The same is true of the average of the
localization operator, which is linear in $\Delta(\mathbf{r},t)$. Therefore,
if we take an average over many realizations of the experiment, and if
$\Delta(\mathbf{r},t)$ and $\rho(t)$ remain uncorrelated, the average
contribution of the localization term in the right-hand side of
(\ref{calc-119}) vanishes.\ It is non-zero only when $\rho(t)$  and the
fluctuations of the Bohmian positions around their quantum equilibrium
positions become correlated.

The situation is therefore similar to a relaxation
phenomenon created by an ensemble of $\mathbf{r}$-dependent fluctuating
perturbations $\Delta(\mathbf{r},t)$. The so called \textquotedblleft
motional narrowing\textquotedblright\ condition (see for instance
\cite{Abragam, vol-2}) expresses that the perturbations have very little effect during
their correlation time. In our case, for a single particle, this condition
reads $\gamma_{L}\tau_{c}\ll1$, which is easily fulfilled with the very small
value (\ref{calc-117}) of $\gamma_{L}$. The same remains obviously true for any
microscopic system: the appearance of weak correlations between the quantum
state of the system and the fluctuations of the Bohmian positions creates a
relaxation process with a rate $\gamma_{L}^{2}\tau_{c}$, which remains negligible over a
time equal to the age of the Universe. We therefore recover the standard
equation of evolution of the density operator.

For a macroscopic system, the situation may be completely different: we have
seen in \S~\ref{various-situations} that the localization term itself grows
quadratically with the number of particles involved, so that the second order
rate of localization $\gamma_{L}^{2}\tau_{c}$ is now multiplied by the fourth
power of the number of particles (assuming that $\tau_c$ is independent of $N$). Since the Avogadro number is very large, one
can easily obtain situation where the rate becomes very fast, and where the
motional narrowing condition is actually no longer valid.\ This corresponds to
situations where the von Neumann projection postulate may be applied and where the measurement apparatuses
can be treated classically.

\subsection{Partial traces}

Assume that the complete system $S$
is made of two subsystems $S_{A}$ and $S_{B}$, which are localized in two
disconnected regions of space $\mathscr{V}_{A}$ and $\mathscr
{V}_{B}$, and
contain $N_{A}$ and $N_{B}$ particles respectively, and have no mutual
interaction.\ The two density functions $N_{A}(\mathbf{r})$ and $N_{B}(\mathbf{r})$ then have non-overlapping supports, so that both the Hamiltonian
and the localization operator are then the sum of two terms:
\begin{subequations}
\label{calc-122}%
\begin{align}
H\left(  t\right)   &  =H_{A}\left(  t\right)  +H_{B}\left(  t\right)
\label{calc-122a}\\
\overline{L}\left(  t\right)   &  =\overline{L}_{A}\left(  t\right)
+\overline{L}_{B}\left(  t\right)  \label{calc-122b}%
\end{align}
In the space of states of a single particle, we choose a basis $\left\{
\left\vert u_{i}\right\rangle \right\}  $ such that each of these states is
localized, either in $\mathscr{V}_{A}$, or $\mathscr{V}_{B}$ (its wave function is zero
in the other volume).\ A basis in the space of states of $S$ can be obtained
with states where the occupation number $n_{i}$ of each $\left\vert
u_{i}\right\rangle $ is specified, that is with the ensemble of kets:%
\end{subequations}
\begin{equation}
\left\vert u_{1}:n_{1};u_{2}:n_{2};...;u_{P}:n_{P}\right\rangle =\left\vert
n_{A},n_{B}\right\rangle \label{calc-122-2}%
\end{equation}
where $n_{A}$ is a condensed notation for all the occupation numbers of the
sates localized in $\mathscr{V}_{A}$, and similarly $n_{B}$ a condensed
notation for the occupation numbers of the states localized in $\mathscr{V}_{B}$. The matrix elements of the density operator $\rho$ describing $S$ are:%
\begin{equation}
\left\langle n_{A},n_{B}\right\vert \rho(t)\left\vert n_{A}^{\prime}%
,n_{B}^{\prime}\right\rangle \label{calc-122-3}%
\end{equation}

Any operator $A$ acting in $\mathscr{V}_{A}$ but not $\mathscr{V}_{B}$ changes
the value of $n_{A}$ but not that of $n_{B}$.\ The average $\left\langle
A\right\rangle $ of $A$ can therefore be obtained from the spatial trace
$\rho_{A}$ of $\rho$ over \ region $\mathscr{V}_{B}$ defined as:%
\begin{equation}
\left\langle n_{A}\right\vert \rho_{A}(t)\left\vert n_{A}^{\prime
}\right\rangle =\sum_{n_{B}}\left\langle n_{A},n_{B}\right\vert \rho
(t)\left\vert n_{A}^{\prime},n_{B}^{}\right\rangle \label{calc-122-4}%
\end{equation}
where the sum over $n_{B}$ is taken over all possible values of the occupation
number of the states localized in $\mathscr{V}_{B}$. The time evolution of
$\rho_{A}(t)$ is obtained by taking the spatial trace of (\ref{calc-119}). The
terms in $H_{A}\left(  t\right)  $ and $\overline{L}_{A}\left(  t\right)  $
give the same effect as in (\ref{calc-119}), with indices $A$ added to the
operators.\ Moreover, as usual the term in $H_{B}\left(  t\right)  $ vanishes
(the partial trace of the commutator is zero).\ As for the term in
$\overline{L}_{B}\left(  t\right)  $, it leads to:%
\begin{align}
\left.  \frac{\text{d}}{\text{d}t}\right\vert _{L_{B}}\left\langle
n_{A}\right\vert \rho_{A}(t)\left\vert n_{A}^{\prime}\right\rangle  &
=\gamma_{L}\sum_{n_{B}^{},n_{A}^{\prime\prime},n_{B}^{\prime\prime}}\left\{
\left\langle n_{A},n_{B}\right\vert \overline{L}_{B}\left(  t\right)
\left\vert n_{A}^{\prime\prime},n_{B}^{\prime\prime}\right\rangle \left\langle
n_{A}^{\prime\prime},n_{B}^{\prime\prime}\right\vert \rho(t)\left\vert
n_{A}^{\prime},n_{B}\right\rangle \right. \nonumber\\
&  ~~~~~~~~~~~~~~~\left.  +\left\langle n_{A},n_{B}\right\vert \rho
(t)\left\vert n_{A}^{\prime\prime},n_{B}^{\prime\prime}\right\rangle
\left\langle n_{A}^{\prime\prime},n_{B}^{\prime\prime}\right\vert \overline
{L}_{B}\left(  t\right)  \left\vert n_{A}^{\prime},n_{B}\right\rangle
\right\}  \label{calc-124}%
\end{align}
In the first term inside the summation, $n_{A}^{\prime\prime}=n_{A}$, while in
the second term $n_{A}^{\prime\prime}=n_{A}^{\prime}$, so that the right hand
side of this equation is equal to:%
\begin{equation}
\gamma_{L}\sum_{n_{B}^{},n_{B}^{\prime\prime}}\left\{  \left\langle
n_{B}\right\vert \overline{L}_{B}\left(  t\right)  \left\vert n_{B}%
^{\prime\prime}\right\rangle \left\langle n_{A},n_{B}^{\prime\prime
}\right\vert \rho(t)\left\vert n_{A}^{\prime},n_{B}\right\rangle +\left\langle
n_{A},n_{B}\right\vert \rho(t)\left\vert n_{A}^{\prime},n_{B}^{\prime\prime
}\right\rangle \left\langle n_{B}^{\prime\prime}\right\vert \overline{L}%
_{B}\left(  t\right)  \left\vert n_{B}\right\rangle \right\}  \label{calc-125}%
\end{equation}
where the two terms become identical as soon as the two dummy variables
$n_{B}$ and $n_{B}^{\prime\prime}$ are interchanged.\ We therefore obtain:%
\begin{equation}
\left.  \frac{\text{d}}{\text{d}t}\right\vert _{L_{B}}\left\langle
n_{A}\right\vert \rho_{A}(t)\left\vert n_{A}^{\prime}\right\rangle
=2\gamma_{L}\sum_{n_{B}^{},n_{B}^{\prime\prime}}\left\langle n_{B}\right\vert
\overline{L}_{B}\left(  t\right)  \left\vert n_{B}^{\prime\prime}\right\rangle
\left\langle n_{A},n_{B}^{\prime\prime}\right\vert \rho(t)\left\vert
n_{A}^{\prime},n_{B}\right\rangle \label{calc-126}%
\end{equation}

(i) If the matrix elements of the density operator of $S$ factorize:%
\begin{equation}
\left\langle n_{A},n_{B}\right\vert \rho(t)\left\vert n_{A}^{\prime}%
,n_{B}^{\prime}\right\rangle =\left\langle n_{A}\right\vert \rho
_{A}(t)\left\vert n_{A}^{\prime}\right\rangle \times\left\langle
n_{B}\right\vert \rho_{B}(t)\left\vert n_{B}^{\prime}\right\rangle
\label{calc-127}%
\end{equation}
we get:%
\begin{equation}
\left.  \frac{\text{d}}{\text{d}t}\right\vert _{L_{B}}\left\langle
n_{A}\right\vert \rho_{A}(t)\left\vert n_{A}^{\prime}\right\rangle
=2\gamma_{L}\left\langle n_{A}\right\vert \rho_{A}(t)\left\vert n_{A}^{\prime
}\right\rangle ~\text{Tr}_{B}\left\{  \overline{L}_{B}\left(  t\right)
\rho_{B}(t)\right\}  \label{calc-128}%
\end{equation}
But we have seen in (\ref{calc-120-2}) that the trace in the right-hand side
vanishes.\ If two subsystems occupy different regions of space, and if they
are uncorrelated, each partial density operator evolves independently (as is
the case in the absence of the localization term).

(ii) If the density operator of $S$ does not factorize, the preceding
simplification does not occur. Let us first study the evolution of the density matrices in a
single realization of an experiment. If the two systems $S_{A}$ and $S_{B}$ are
entangled, the dynamical collapse acting on $S_{B}$ may affect the state of
$S_{A}$, in the same way as the standard von Neumann collapse postulate can
change at the same time the state of two remote entangled systems. Mathematically,
the origin of this mutual effect of the two subsystems is the anticommutator
that contains $\overline{L}\left(  t\right)  $ in (\ref{calc-119}), while the
Hamiltonian appears in a commutator.\ Therefore, in the partial trace, while the two
terms in $H_{B}$ cancel each other, the two terms in $\overline{L}_B\left(
t\right)  $ add to provide the double of each contribution.

The quantum nonlocality then manifests itself in two ways.\ The first
also occurs in standard dBB theory, where the motion of the Bohmian positions
of the whole system are guided in the configuration space by the wave function
in this space.\ The second is due to the nonlocal effect of the collapse term
in the equation of evolution. This effect is necessary to recover the results provided by the usual
von Neumann reduction postulate in standard quantum mechanics.

(iii) Nevertheless, if many realizations of the experiment are performed, one has to consider the average of the localization operator $\overline{L}_{B}$ over these realizations. We have discussed in \S~\ref{quantum-equilibrium} the conditions under which the quantum equilibrium is obtained. If this is the case, $N_{\Phi}\left(  \mathbf{r},t\right)  $ and $N_{\rho}\left(\mathbf{r},t\right)$ are constantly equal, and the average of $\overline{L}_{B}$ vanishes. Therefore, there can be no influence of an experiment performed in region $B$ on the density operator in region $A$, which  automatically ensures the
non-signaling property necessary to obtain a model that is compatible with relativity. We recover the relation obtained by Valentini \cite{Valentini-2002} between quantum equilibrium and the no-signaling condition, which is thus also valid within our non-standard model.

\section{Effect of the localization term on the densities and currents}
\label{position-representation}

We now study the effect of the localization term on the density of particles and on their current. As before, for the sake of simplicity, we assume that the particles are spinless.

\subsection{Evolution of the one-body density}
\label{one-body-density}

We begin with the study of a pure state. The wave function is symmetric with respect to the exchange of particles. The
one body density is then:%
\begin{equation}
D_{\Phi}\left(  \mathbf{r}\right)  =N\int\text{d}^{3}r_{2}...\text{d}^{3}%
r_{N}~\left\vert \overline{\Phi}\left(  \mathbf{r}_{1}=\mathbf{r}%
,\mathbf{r}_{2},...,\mathbf{r}_{N};t\right)  \right\vert ^{2} \label{calc-5}%
\end{equation}
According to (\ref{calc-4}), the contribution of the localization term to its time evolution is given by:%
\begin{equation}
\left.  \frac{\text{d}}{\text{d}t}\right\vert _{\text{loc}}D_{\Phi}\left(
\mathbf{r}\right)  =2\gamma_{L}N\int\text{d}^{3}r_{2}...\text{d}^{3}%
r_{N}~\left[  \sum_{n=1}^{N}\Delta\left(  \mathbf{r}_{n}\mathbf{,}t\right)
-N<\Delta>\right]  \left\vert \overline{\Phi}\left(  \mathbf{r}_{1}%
=\mathbf{r},\mathbf{r}_{2},..,\mathbf{r}_{n},..,\mathbf{r}_{N};t\right)
\right\vert ^{2} \label{calc-101}%
\end{equation}

In (\ref{calc-101}), the term $n=1$ merely introduces a term proportional to
$\Delta\left(  \mathbf{r,}t\right)  D_{\Phi}\left(  \mathbf{r}\right)  $,
which depends only on the single particle density; all the other terms contain
the position correlation function $D_{\Phi}^{II}\left(  \mathbf{r}%
,\mathbf{r}^{\prime}\right)  $ of two particles at points $\mathbf{r}$ and
$\mathbf{r}_{p}$:%
\begin{equation}
D_{\Phi}^{II}\left(  \mathbf{r},\mathbf{r}^{\prime}\right)  =N\left(
N-1\right)  \int\text{d}^{3}r_{3}...\text{d}^{3}r_{N}~\left\vert
\overline{\Phi}\left(  \mathbf{r}_{1}=\mathbf{r},\mathbf{r}_{2}=\mathbf{r}%
^{\prime},...,\mathbf{r}_{N};t\right)  \right\vert ^{2} \label{calc-201}%
\end{equation}
Equation (\ref{calc-101}) \ then provides:%
\begin{equation}
\left.  \frac{\text{d}}{\text{d}t}\right\vert _{\text{loc}}D_{\Phi}\left(
\mathbf{r}\right)  =2\gamma_{L}\left[  \Delta\left(  \mathbf{r,}t\right)
~D_{\Phi}\left(  \mathbf{r}\right)  +\int\text{d}^{3}r^{\prime}~D_{\Phi}%
^{II}\left(  \mathbf{r},\mathbf{r}^{\prime}\right)  ~\Delta\left(
\mathbf{r}^{\prime}\mathbf{,}t\right)  -N<\Delta>D_{\Phi}\left(
\mathbf{r}\right)  \right]  \label{calc-202}%
\end{equation}
We do not get a closed equation for the evolution of the single particle
density; the right hand side of this equation contains the two-particle
density, as in the usual BBGKY\ hierarchy, and despite of the fact that the
localization term is a single-particle operator.\ This is because the
localization term is non-Hermitian, which introduces anticommutators instead
of commutators.

Since:%
\begin{equation}
\int\text{d}^{3}r~D_{\Phi}^{II}\left(  \mathbf{r},\mathbf{r}^{\prime}\right)
=\left(  N-1\right)  ~D_{\Phi}\left(  \mathbf{r}^{\prime}\right)
\label{calc-103}%
\end{equation}
we can check the particle conservation rule:%
\begin{align}
\int\text{d}^{3}r\left.  \frac{\text{d}}{\text{d}t}\right\vert _{\text{loc}%
}D_{\Phi}\left(  \mathbf{r}\right)   &  =2\gamma_{L}\left[  \int\text{d}%
^{3}r~\Delta\left(  \mathbf{r,}t\right)  ~D_{\Phi}\left(  \mathbf{r}\right)
+\left(  N-1\right)  \int\text{d}^{3}r^{\prime}~D_{\Phi}\left(  \mathbf{r}%
^{\prime}\right)  ~\Delta\left(  \mathbf{r}^{\prime}\mathbf{,}t\right)
-N^{2}<\Delta>\right] \nonumber\\
&  =2\gamma_{L}\left[  N\int\text{d}^{3}r~\Delta\left(  \mathbf{r,}t\right)
~D_{\Phi}\left(  \mathbf{r}\right)  -N^{2}<\Delta>\right]  =0 \label{calc-104}%
\end{align}

If the system is not described by a pure state, but by a density
operator $\rho$, the time evolution of the density $D_{\rho}(\mathbf{r},t)$ is
obtained by the same calculation, with the simple substitution:%
\begin{equation}
\left\vert \overline{\Phi}\left(  \mathbf{r}_{1}=\mathbf{r},\mathbf{r}%
_{2},...,\mathbf{r}_{N};t\right)  \right\vert ^{2}\Rightarrow\left\langle
\mathbf{r}_{1}=\mathbf{r},\mathbf{r}_{2},...,\mathbf{r}_{N}\right\vert
\rho(t)\left\vert \mathbf{r}_{1}=\mathbf{r},\mathbf{r}_{2},...,\mathbf{r}%
_{N}\right\rangle \label{calc-120-3}%
\end{equation}
in all the equations, including (\ref{calc-201}), which becomes the definition
of the two body density $D_{\rho}^{II}\left(  \mathbf{r},\mathbf{r}^{\prime
}\right)  $. The time evolution of $D_{\rho}(\mathbf{r},t)$ is therefore given
by:%
\begin{equation}
\left.  \frac{\text{d}}{\text{d}t}\right\vert _{\text{loc}}D_{\rho}\left(
\mathbf{r}\right)  =2\hslash\gamma_{L}\left[  \Delta\left(  \mathbf{r,}%
t\right)  ~D_{\rho}\left(  \mathbf{r}\right)  +\int\text{d}^{3}r^{\prime
}~D_{\rho}^{II}\left(  \mathbf{r},\mathbf{r}^{\prime}\right)  ~\Delta\left(
\mathbf{r}^{\prime}\mathbf{,}t\right)  -N<\Delta>D_{\rho}\left(
\mathbf{r}\right)  \right]  \label{calc-202-bis}%
\end{equation}

\subsection{Mean field, role of the spatial correlations}

When the distance between $\mathbf{r}$ and $\mathbf{r}^{\prime}$ becomes very
large, the correlation function $D_{\Phi}^{II}\left(  \mathbf{r}%
,\mathbf{r}^{\prime}\right)  $ factorizes.\ We can therefore introduce a
function $F(\mathbf{r},\mathbf{r}^{\prime})$ by setting:%
\begin{equation}
D_{\Phi}^{II}\left(  \mathbf{r},\mathbf{r}^{\prime}\right)  =~D_{\Phi}\left(
\mathbf{r}\right)  ~D_{\Phi}\left(  \mathbf{r}^{\prime}\right)  \left[
1-F(\mathbf{r},\mathbf{r}^{\prime})\right]  \label{A4}%
\end{equation}
with:%
\begin{equation}
F(\mathbf{r},\mathbf{r}^{\prime})\underset{\left\vert \mathbf{r}%
-\mathbf{r}^{\prime}\right\vert \rightarrow\infty}{\rightarrow}0 \label{A7}%
\end{equation}
In (\ref{A4}) we have chosen to write a minus sign before
$F(\mathbf{r},\mathbf{r}^{\prime})$ because, if two neighbor systems exchange
particles, their local densities are anticorrelated; $F$ is then
positive.\ Relation (\ref{calc-103}) provides:%
\begin{equation}
\int\text{d}^{3}r^{\prime}~D_{\Phi}\left(  \mathbf{r}^{\prime}\right)
~F(\mathbf{r},\mathbf{r}^{\prime})=1 \label{A6}%
\end{equation}
The function $F(\mathbf{r},\mathbf{r}^{\prime})$ may change sign in general,
but positive values dominate this integral.

If we insert (\ref{A4})\ into (\ref{calc-202}), the first term in the right
hand side cancels the term in $N<\Delta> D_{\Phi}\left(
\mathbf{r}\right)$. Using (\ref{A6}), we
get:%
\begin{align}
\left.  \frac{\text{d}}{\text{d}t}\right\vert _{\text{loc}}D_{\Phi}\left(
\mathbf{r}\right)   &  =2\gamma_{L}\left[  \Delta\left( \mathbf{r,}t\right)
-\int\text{d}^{3}r^{\prime}~F(\mathbf{r},\mathbf{r}^{\prime})~D_{\Phi}\left(
\mathbf{r}^{\prime}\right)  ~\Delta\left(\mathbf{r}^{\prime}\mathbf{,} t\right)  \right]  D_{\Phi}\left(  \mathbf{r}\right) \nonumber\\
&  =2\gamma_{L}D_{\Phi}\left(  \mathbf{r}\right)  \int\text{d}^{3}r^{\prime
}~F(\mathbf{r},\mathbf{r}^{\prime})~D_{\Phi}\left(  \mathbf{r}^{\prime
}\right)  ~\left[  \Delta\left(  \mathbf{r,}t\right)  -\Delta\left(
\mathbf{r}^{\prime}\mathbf{,}t\right)  \right]  \label{A8}%
\end{align}
The \textquotedblleft local\textquotedblright\ character of this equation of
evolution depends on the properties of $F(\mathbf{r},\mathbf{r}^{\prime})$, in
particular whether it tends to zero sufficiently rapidly when the difference
of positions increases.

(i) In mean-field theory, one merely assumes that  $F(\mathbf{r},\mathbf{r}^{\prime})$ vanishes. The first line of relation (\ref{A8}) then becomes:
\begin{equation}
\left.  \frac{\text{d}}{\text{d}t}\right\vert _{\text{loc}}D_{\Phi}\left(\mathbf{r}\right)    =2\gamma_{L} \Delta\left( \mathbf{r,}t\right) D_{\Phi}\left(  \mathbf{r}\right)
\label{chp-moyen}
\end{equation}
Mean field theory merely predicts that $D_{\Phi}\left(\mathbf{r}\right)$ increases at points where $\Delta\left( \mathbf{r,}t\right)$ is positive, decreases at points where this function is negative.

(ii) Beyond mean-field theory, if $F(\mathbf{r},\mathbf{r}^{\prime})$ has a small range $l$ (range of
correlations in the system), the localization term depends only on the values
of $\Delta\left(  \mathbf{r}^{\prime}\mathbf{,}t\right)  $ in a small domain
around $\mathbf{r}$. In the limit of a very small range where:%
\begin{equation}
F(\mathbf{r},\mathbf{r}^{\prime})\sim\delta(\mathbf{r}-\mathbf{r}^{\prime})
\label{A10}%
\end{equation}
The right-hand side of (\ref{A8}) vanishes.\ More generally, if $\Delta\left(
\mathbf{r}^{\prime}\mathbf{,}t\right)  $ is constant in the domain where
$F(\mathbf{r},\mathbf{r}^{\prime})$ is not zero, relation (\ref{A8}) shows
that the evolution of the density introduced by the localization process
vanishes. If $\Delta\left(  \mathbf{r}^{\prime}\mathbf{,}t\right)  $ varies in
space over a distance $l$, the integral can approximated by:%
\begin{equation}
- \int_{r^{\prime}\lesssim l}\text{d}^{3}r^{\prime} ~ F(\mathbf{r},\mathbf{r}%
^{\prime})~D_{\Phi}\left(  \mathbf{r}^{\prime}\right)  ~(\mathbf{r^{\prime}%
}-\mathbf{r})\cdot\bm{\nabla}\Delta\left(  \mathbf{r,}t\right)
\label{calc-111-2}%
\end{equation}
If the product $F(\mathbf{r},\mathbf{r}^{\prime})D_{\Phi}\left(
\mathbf{r}^{\prime}\right)  $ varies linearly as a function of $\mathbf{r}%
^{\prime}$ within its range $l$:%
\begin{equation}
F(\mathbf{r},\mathbf{r}^{\prime})~D_{\Phi}\left(  \mathbf{r}^{\prime}\right)
=F(\mathbf{r},\mathbf{0})~D_{\Phi}\left(  \mathbf{0}\right)
+(\mathbf{r^{\prime}}-\mathbf{r})\cdot\bm{\nabla}(FD) \label{calc-200}%
\end{equation}
we obtain:%
\begin{equation}
\left.  \frac{\text{d}}{\text{d}t}\right\vert _{\text{loc}}D_{\Phi}\left(
\mathbf{r}\right)  =-\gamma_{L}\frac{4\pi l^{5}}{15}\bm{\nabla}(FD)\cdot
\bm{\nabla}\Delta\left(  \mathbf{r,}t\right)  \label{calc-401}%
\end{equation}
This expression varies very rapidly with the range of correlation $l$.

(iii) But $F(\mathbf{r},\mathbf{r}^{\prime})$ can also have a large range. For
instance, just after a measurement has been performed, the correlation
function between two different positions of the pointer vanishes, while the
produce of one-body densities does not.\ In this case, the transfer of density due to the localization term
is not local. This is a necessary feature to eliminate MDQS
efficiently, and to obtain a projection after measurement.

We conclude from this discussion that the localization term has little effect
in most (ordinary) situations.\ But, if a MDQS\ appears for some
reason (Schr\"{o}dinger cat, etc.), it is promptly reduced to one of its
components by the localization term.

\subsection{Evolution of the local momentum}
\label{local-momentum}

We have only studied the direct effects of the localization term on the
local density, but of course this term also has indirect effects: by
localizing the wave functions in space, it changes their Fourier transform,
and therefore the average value of the velocities.\ At later times, this
change will also modify the average positions of the particles.

The local current of particles at point $\mathbf{r}$ is:%
\begin{align}
\mathbf{J}_{\Phi}(\mathbf{r},t)=\frac{\hbar}{2im}%
{\displaystyle\sum\limits_{p=1}^{N}}
\int\text{d}^{3}r_{1}  &  ...\int\text{d}^{3}r_{p-1}\int\text{d}^{3}%
r_{p+1}...\int\text{d}^{3}r_{N}\nonumber\\
&  \overline{\Phi}^{\;\ast}\left(  \mathbf{r}_{1},\mathbf{r}_{2}%
,...,\mathbf{r}_{p}=\mathbf{r},...\mathbf{r}_{N};t\right)
~\bm{\nabla}_{\mathbf{r}}~\overline{\Phi}\left(  \mathbf{r}_{1},\mathbf{r}%
_{2},...,\mathbf{r}_{p}=\mathbf{r},...\mathbf{r}_{N};t\right)  ~+\text{c.c.}
\label{abd-1}%
\end{align}
where c.c. means complex conjugate. The time derivative of this current
induced by the localization process is obtained by using equation
(\ref{calc-4}).\ The  derivative of $\overline{\Phi}\left(  \mathbf{r}%
_{1},\mathbf{r}_{2},...,\mathbf{r}_{p}=\mathbf{r},...\mathbf{r}_{N};t\right)
$ introduces the expression:%
\begin{align}
 \frac{\hbar \gamma_{L}}{2im} {\displaystyle\sum\limits_{p=1}^{N}} \bm{\nabla}_{\mathbf{r}}~\Big[  \Delta
(\mathbf{r},t) & +\sum_{n\neq p} \Delta(\mathbf{r}_{n},t)-N\Delta\Big]  \overline{\Phi}\left(
\mathbf{r}_{1},\mathbf{r}_{2},...,\mathbf{r}_{p}=\mathbf{r},...\mathbf{r}%
_{N};t\right) \nonumber\\
&  \frac{\hbar \gamma_{L}}{2im} \Big[  \Delta(\mathbf{r},t)+ \sum_{n\neq p} \Delta(\mathbf{r}%
_{n},t)-N\Delta \Big]  \overline{\Phi}\bm{\nabla}_{\mathbf{r}}~\overline
{\Phi}\left(  \mathbf{r}_{1},\mathbf{r}_{2},...,\mathbf{r}_{p}=\mathbf{r}%
,...\mathbf{r}_{N};t\right) \nonumber\\
& \hspace{3cm}+ \frac{\hbar \gamma_{L}}{2im}  \left[  \bm{\nabla}_{\mathbf{r}}%
~\Delta(\mathbf{r},t)\right]  \overline{\Phi}\left(  \mathbf{r}_{1}%
,\mathbf{r}_{2},...,\mathbf{r}_{p}=\mathbf{r},...\mathbf{r}_{N};t\right)
\label{abd-2}%
\end{align}
The first term in the right hand side of the second line
reconstructs the particle current $\mathbf{J}(\mathbf{r},t)$, multiplied by
$\Delta(\mathbf{r},t)$; the rest of the second line introduces a new integral.
The term in the third line introduces the single particle density
(\ref{calc-5}), which is real; since the whole term is multiplied by $i$ in
(\ref{abd-1}), this term disappears when the real part is taken.\ If we
combine these terms with those resulting from the derivative of $\overline
{\Phi}^{\;\ast}$, we obtain:%
\begin{align}
\left.  \frac{\text{d}}{\text{d}t}\right\vert _{\text{loc}}\mathbf{J}_{\Phi
}\left(  \mathbf{r}\right)   &  =2\gamma_{L}\left[  \Delta(\mathbf{r}%
,t)-\Delta\right]  \mathbf{J}_{\Phi}\left(  \mathbf{r}\right) \nonumber\\
&  +\frac{\hbar\gamma_{L}}{2im}%
{\displaystyle\sum\limits_{p=1}^{N}}
\int\text{d}^{3}r_{1}...\int\text{d}^{3}r_{p-1}\int\text{d}^{3}r_{p+1}%
...\int\text{d}^{3}r_{N}\sum_{n\neq p}\left[  \Delta(\mathbf{r}_{n}%
,t)-\Delta\right]  ~\overline{\Phi}^{\;\ast}\bm{\nabla}_{\mathbf{r}}%
~\overline{\Phi}~+\text{c.c.} \label{abd-3}%
\end{align}
The first line of this equation adds an exponential increase or decrease of
the particle current with a fluctuating rate, which can be positive or
negative. The second line couples the current $\mathbf{J}_{\Phi}\left(
\mathbf{r}\right)  $ to another integral. Since $\gamma_{L}$ is very small,
the influence of the localization process on the current of particles remains very
small in most situations, except when the difference $\left[
\Delta(\mathbf{r}_{n},t)-\Delta\right]  $ can take on very large values, as is the case during  a measurement process (\S~\ref{various-situations}).

In standard GRW or CSL theory, the random localization process is a pure Markov process. The corresponding changes of the momentum of an object are then given by a random walk with no memory, and no preferred direction \cite{Collett-Pearle-2003}. In the model of the present article, the localization process has a non-zero memory arising from the statistical properties of $\Delta(\mathbf{r},t)$. These properties depend on the complicated nonlinear relative motion of the Bohmian positions and the wave function. Of course, for macroscopic objects, the corresponding time constants are very large, due to the very small value of the localization time constant $\gamma_L$. Nevertheless, on very long time scales such as those often considered in astrophysics, it may be that observable effects are predicted. Such predictions are nevertheless difficult to make, since it is not easy to evaluate the spatial and temporal scales of the fluctuations of the localization source $\Delta (\mathbf r ,t )$.

\section{Possible interpretations}

In terms of possible interpretations of quantum mechanics, the model is relatively robust: it can remain compatible with very different points of view.\

One can consider that the Bohmian variables are just a mathematical
tool to introduce the stochastic reduction of the state vector, which then
represents physical reality, in the line of GRW and CSL\ theories.\ Indeed, we have not assumed that Bohmian variables directly provide the results of position
measurements, but that the results are determined by the quantum density in space provided by the state vector: they
 depend on the values of
$D_{\Phi}\left(  \mathbf{r}\right)  $, not on those of the Bohmian
positions. One may also consider that $D_{\Phi}\left(  \mathbf{r}\right)  $ gives a direct description of physical reality or ordinary 3D space, coming back to a fluid representation of matter, as envisaged initially by
Schr\"{o}dinger when he introduced his equation. This view is just the opposite of the usual dBB theory, where it is assumed that the observations reveal the values of the
Bohmian positions, which therefore directly represent the physical reality.
Here, the position in the configuration space is just a mathematical variable that drives the wave function and the associated quantum density in space; it plays a role that is analogous to the \textquotedblleft
subquantum medium\textquotedblright\ acting on the state vector in
F\'{e}nyes-Nelson theories \cite{Fenyes, Nelson}. One can even
combine this new dynamics with the Everett interpretation; if one includes the memory registers of the observers into $\Psi$,
one obtains a sort of \textquotedblleft Everett interpretation with
projection\textquotedblright, predicting the existence of a \textquotedblleft
single world\textquotedblright.

But one
can also prefer the usual approach of the dBB theory, and consider that
the Bohmian positions represent the beables \cite{Bell-local-beables, Bell-livre} of the
physical system.\ The advantage of the model is then to get rid of all the
empty waves of the dBB theory, and of the relative difficulty to attribute
them a status \cite{Lewis-2007}. If one sees the wave function as similar to
a Lagrangian or Hamiltonian in classical mechanics \cite{Bricmont}, it seems preferable to keep only the effective part of this wave function,
eliminating all empty waves that have accumulated in the past.

One can also
take an intermediate point of view, and consider for instance that what
represents physical reality is the spatial density $N_{B}(\mathbf{r},t)$.
 In this view, the problem of the \textquotedblleft long tails\textquotedblright\ occurring in the usual spontaneous theories becomes irrelevant, since no particular physical meaning is attributed to the exponentially vanishing empty waves.

Whatever status is eventually attributed to the state vector $\Psi$, it remains clear that it
is less disconnected from physical reality than with a dynamics having no
collapse mechanism. Nevertheless, the various
mathematical components of this dynamics can be interpreted in different ways,
leading to various ontologies.

\section{Discussion and conclusion}

We have seen that the attraction of the Bohmian densities can be used to obtain a reasonable
model of spontaneous collapse of the state vector. In addition to the standard \textquotedblleft pilot wave\textquotedblright\ of the dBB theory, we have introduced a \textquotedblleft pilot density\textquotedblright\ for the wave. The corresponding dynamic is
deterministic: the stochasticity of the initial position in configuration space is sufficient
to reproduce the standard prediction of the Born rule for probabilities. The
localization term tends to constantly adapt the wave function in order to
obtain a better match between the quantum density
 and the density of positions in space. In QSMDS situations, all the empty waves disappear, so that the wave function and the Bohmian positions progress together in time, while in usual dBB theory they are disconnected. For instance, it has been claimed that the dBB theory is really \textquotedblleft a many-world theory with a superfluous configuration appended to one of the worlds\textquotedblright , and that \textquotedblleft pilot-wave theories are parallel-universes theories in a state of chronic denial  \textquotedblright\ \cite{Deutsch-1986}; for discussions of these claims, see for instance  \cite{Valentini-2007} and \cite{Brown-2009}. Clearly, if one accepts the present model where the dynamics of the state vector is coupled to the Bohmian positions, this discussion is settled.

The mechanism of the projection in the dynamics
is actually based on the cohesion of macroscopic objects.\ Because standard theory
predicts that, even if such objects reach in QSMDS, their constituent
particles remain strongly correlated spatially, the mechanism of our model
projects them into a single localization.\ In fact, only macroscopic objects
that do not break spontaneously into several parts acquire in this way a unique
spatial localization: the \textquotedblleft moon is there even if
nobody looks\textquotedblright\ \cite{Mermin}, and the reason why the center
of the moon occupies a well defined point on its orbit before any measurement
is the internal cohesion of the moon. By contrast, objects
than can be split into several components without breaking any energy barrier can go trough QSMDS without collapsing; for instance,  Bose-Einstein gaseous condensates that are split into two remote parts can give rise later to quantum interference effects, as discussed in more detail in \cite{SDAP}.

The localization term also provides a sharp transition between the quantum
and classical regime.\ For a molecule or cluster containing $N$ particles in a
volume smaller than $a_{L}$, any superposition of several quantum states
localized at a distance larger that $a_{L}$ is projected into one single
component of this superposition with a rate $\Gamma$ that varies quadratically
as a function of $N$:
\begin{equation}
\Gamma\simeq\gamma_{L}~N^{2} \label{gamma-1}%
\end{equation}
With the choice (\ref{calc-112-other}) of constants, and if the number of particles is about $10^{12}$, this relation predicts a
localization time of $1$ s. Since the number density $n$ of solid (or liquid) physical objects is of the order of $10^{30}$ atoms/cubic meter, the same relation
can be expressed in terms of the size of the object:%
\begin{equation}
\Gamma\simeq\gamma_{L}~n^{2}l^{6} \label{gamma-2}%
\end{equation}
For an object of size $l<a_{L}$ containing $n$ particles per unit volume, this
rate varies proportionally to the sixth power of the size $l$; the
localization time is $1$ second if $l\simeq1\mu  $m. For an object of size $l>a_{L}$
containing $N$ particles, the rate becomes:%
\begin{equation}
\Gamma\simeq\gamma_{L}~na_{L}^{3} n l^3 \label{gamma-3}%
\end{equation}
One can therefore consider that $l=a_{L}$ and $N=10^{12}$ provide the border
between standard quantum and classical behavior of physical objects.

It is clear that the model differs from the GRW\ and CSL\ theories in several
respects. Beyond the fact that the dynamics is not stochastic, already mentioned in the introduction, another difference is that the localization process is no longer a single particle process, where each of them is localized independently, but results from a collective effect between the particles; this creates correlations between them, as remarked at the end of \S~\ref{effect-localization}. This collective character introduces  a sharper transition
between the quantum and classical regimes, due to the
quadratic term in $N$ in (\ref{gamma-1}). This in turn is a consequence of the
fact that the number of particles enters twice in the model, once in the
Bohmian number $N_{B}\left(  \mathbf{r},t\right)  $, and once in the integral
of the quantum operator $\Psi^{\dagger}\left(  \mathbf{r}\right)  \Psi\left(
\mathbf{r}\right)  $. In QSMDS\ situations, our localization term is therefore \textquotedblleft stronger\textquotedblright\
than those of GRW\ and CSL.\ Nevertheless, in usual situations, it is \textquotedblleft
softer\textquotedblright: for instance, in a bulk piece of matter located in a
single region of space, the Bohmian density coincides almost perfectly with
$D_{\Phi}\left(  \mathbf{r}\right)  $, and in our model the collapse term has
practically no effect.\ By contrast, it is constantly active in GRW and
CSL\ theories.\ In other words, in experiments such that of Ref
\cite{experiment-heating}, our model would be compatible with the observation
of zero heating.\ We also note that the term that we added to the standard
Schr\"{o}dinger equation is not only non-linear, but also non-local: when the
two components of the wave function of a macroscopic object begin to separate
in space, one component is transferred in space to the other at some finite
distance. This feature is necessary to fit with the quantum predictions in Bell type experiments.

For macroscopic systems, the appearance of position uniqueness is not necessarily the only effect
predicted by the model. We have seen in
\S \ \ref{local-momentum} that another effect of the localization term may be to change the local current of the
particles, which will have an indirect effect on the the density. This is not surprising
since a space localization of the wave function implies an increase of the
width of its Fourier transform, which corresponds to higher velocities. This violates the usual momentum conservation rule; similarly, the GRW and CSL theories predict a spontaneous heating effect that seems to violate the energy conservation rule \cite{Pearle-Mullin-Laloe}.
Whether or not  our model predict slow changes  of the momentum
of macroscopic  objects, for instance on a cosmological time scale,
remains   to be studied.

We have also seen in \S~\ref{quantum-equilibrium} that, within this model, the Born rule is no longer a postulate; it emerges from the dynamics, and it is actually not an exact rule. It nevertheless remains a fantastically good approximation, with an accuracy better than $10^{-16}$ if the preparation and the measurement of quantum system are separated by more than 1s.

Needless to say, the class of models we have discussed is, in a sense, very
naive. It is neither relativistic nor expressed in terms of a plausible field
theory. Its purpose is just to indicate a range of possibilities, which might be exploited  in a second stage
to construct a more credible theory. This range is relatively broad, for  several  reasons. The first obvious reason is that there exist a large domain of possible values for
the two constants $a_{L}$ and $\gamma_{L}$ that introduce no contradiction
with the known experimental facts; this robustness of the model is both a
strength and a weakness, since having too much flexibility amounts to reducing
the predictive power of a theory. A second reason is that the equations are not particularly plausible: they are just the simplest version of a dynamics where position variables are added to the state vector, and where this vector is attracted towards these additional positions. The localization function
$A_{L}(\mathbf{r}-\mathbf{r}^{\prime}\mathbf{)}$ that we have introduced in
(\ref{sdap-2-bis}) is arbitrary, and the choice of a Gaussian has no
particular justification.\ Even the form of the operator $L\left(  t\right)  $
could be changed: it might be possible to re-introduce at this stage the
effect of gravitation by choosing another localization operator, more in the
line of the ideas of Refs.\ \cite{Diosi-1989, GGR-1990, Penrose-1996}; such a
possibility remains to be explored. Finally, we have not specified the
nature of the particles on which the localization process should apply: they could be for
instance, nucleons only, or nucleons and electrons, or quarks, etc. In any
case, at this stage, even the order of magnitude of the constants is
not determined. As mentioned in the introduction, the main merit of this class of models is only to show that they various quantum descriptions of physical reality remain compatible with known experimental data.

\end{document}